\documentclass[
,english
,aip
,amsmath
,amssymb
,reprint
]{revtex4-1}

\usepackage[utf8]{inputenc}
\usepackage[T1]{fontenc}
\usepackage{xcolor}
\usepackage{float}
\usepackage{bm}
\usepackage{graphicx}
\usepackage[version=3]{mhchem}
\usepackage{babel}
\usepackage{dcolumn}

\preprint{AIP/123-QED}

\newcommand{\cmr}{~\mathrm{cm}^{-1}}

\renewcommand{\bm}{\mathbf}

\begin{document}

\title{
Machine Learning for Vibrational Spectroscopy via Divide-and-Conquer
Semiclassical Initial Value Representation Molecular Dynamics with
Application to N-Methylacetamide
}

\author{Michele Gandolfi}
\author{Alessandro Rognoni}
\author{Chiara Aieta}
\author{Riccardo Conte}
\author{Michele Ceotto}
\email{michele.ceotto@unimi.it}
\affiliation{Dipartimento di Chimica, Università degli Studi di Milano, via Golgi
19, 20133 Milano, Italy}

\date{\today}
\homepage{https://sites.unimi.it/ceotto/}

\begin{abstract}
A machine learning algorithm for partitioning the nuclear vibrational
space into subspaces is introduced. The subdivision criterion is based
on Liouville's theorem, i.e. best preservation of the unitary of the
reduced dimensionality Jacobian determinant within each subspace along
a probe full-dimensional classical trajectory. The algorithm is based
on the idea of evolutionary selection and it is implemented through
a probability graph representation of the vibrational space partitioning.
We interface this customized version of genetic algorithms
with our divide-and-conquer semiclassical initial value representation
method for calculation of molecular power spectra. First, we benchmark
the algorithm by calculating the vibrational power spectra of two
model systems, for which the exact subspace division is known. Then,
we apply it to the calculation of the power spectrum of methane. Exact
calculations and full-dimensional semiclassical spectra of this small
molecule are available and provide an additional test of the accuracy
of the new approach. Finally, the algorithm is applied to the divide-and-conquer
semiclassical calculation of the power spectrum of 12-atom \emph{trans}-N-Methylacetamide.
\end{abstract}

\maketitle

\section{Introduction}

When full-dimensional calculations are not computationally feasible,
one needs to introduce some sort of approximation. This happens quite
often in quantum molecular dynamics because the full-dimensional potential
V(\textbf{q}) is generally non separable. However, if the system potential
were made of the sum of a finite number of lower dimensional terms
of the type $V_{a}\left(q_{1},\:q_{2}\right)+V_{b}\left(q_{\text{3}}\right)+V_{c}\left(q_{4},\:q_{5},\:q_{6}\right)+...$,
then each collection of variables within the several terms would compose
a sensible subspace suitable for an independent calculation. Unfortunately,
this is not the case for nuclear dynamics, but one may wonder which
is the decomposition of the full-dimensional vibrational space into
subspaces such that the couplings between different subspace modes
are minimized. The main goal of this paper is to provide a possible
answer to this issue by means of a clustering optimization algorithm.

The algorithm we propose has its roots in the sound ground of Genetic
Algorithms (GAs), but it is different for technical definitions and
implementation and we will refer to it with the more general label
of Evolutionary Algorithm. The theoretical background and implementation
of GAs could be traced back to the works of Holland, Goldberg, and
Henry.\cite{holland1962outline,goldberg1988genetic,holland1992adaptation}
Since then, they have been successfully applied to various problems
in analytical chemistry and chemometrics, such as the analysis of
NMR pulse patterns from complex molecules,\cite{freeman1987design}
the optimal choice of wavelengths to determine the concentration of
a sample, and conformational analysis. \cite{hibbert1993genetic}
GAs have been proven to be the first choice in many cases of feature
selection in regression and classification problems in general \cite{leardi2001genetic,niazi2012genetic}
and in Quantitative Structure Activity Relationships in particular.\cite{beheshti2016qsar,niazi2012genetic}
Evolutionary Algorithms have also been used in materials science to
study the dynamical properties of molecules on surfaces during molecular
dynamics simulations.\cite{bhattacharya2009analyzing}

We test the accuracy of the new optimization algorithm when applied
to the calculation of vibrational power spectra using semiclassical
molecular dynamics. Specifically, we employ the divide-and-conquer
approach to the time-averaged semiclassical initial value representation
(DC-SCIVR) method for nuclear power spectrum calculations.\cite{ceotto_conte_DCSCIVR_2017}
In the DC-SCIVR strategy, the vibrational space is divided into subspaces
to overcome the issue of the full-dimensional calculation. So far,
at the best of our knowledge,
three methods for the partition of the full-dimensional vibrational
problem into lower dimensional subspaces have been proposed. These
are the Hessian method,\cite{DiLiberto_Ceotto_Jacobiano_2018} the 
Wehrle-Šulc-Van\'i\u cek
method,\cite{Wehrle_Vanicek_Oligothiophenes_2014} and the Jacobi
method.\cite{DiLiberto_Ceotto_Jacobiano_2018} According to the latter, the residual
coupling between subspaces is estimated by recurring to Liouville's
theorem. In few words, the best clustering of vibrational modes is
the one where each subspace preserves as much as possible its phase
space volume during a classical, full-dimensional trajectory. This
approach has the advantage of being based on dynamics, i.e. it depends
on both nuclear kinetic and potential contributions. Originally, we
applied this method by searching over all possible subspace combinations.
As a matter of fact, the problem scales approximately as a binomial
factor times the number of subspaces $n_{s}$, i.e. $\sim n_{s}\times F!/D!(F-D)!$,
where $F$ is the number of degrees of freedom and $D$ is the size
of a subspace. Furthermore, it is not possible to know in advance
how many subspaces one should choose and the size of each one.

In this work we introduce an Evolutionary Algorithm and another simplified
and less demanding approach. Both are able to automatize these choices
and highly reduce the computational cost of the optimization. The
paper is organized as follows: In Sec. \ref{sec:Methods}, we recap
the DC-SCIVR method and present the optimization algorithms. Sec.
\ref{sec:Results} presents our results for spectroscopic calculations
and Sec. \ref{sec:Conclusions} concludes the paper and offers some
perspectives.

\section{Methods\label{sec:Methods}}

\subsection{Semiclassical spectra\label{subsec:Semiclassical-spectra}}

The semiclassical power spectrum $I(E)$ of a system described by
the Hamiltonian $\hat{H}$ is equal to the Fourier-transformed wavepacket
survival amplitude (in atomic units)\cite{Heller_SCspectroscopy_1981}
\begin{equation}
I(E)=\frac{1}{2\pi}\intop_{-\infty}^{+\infty}e^{iEt}\langle\chi|\chi(t)\rangle dt,\label{eq:Survival}
\end{equation}
where $|\chi\left(t\right)\rangle=e^{-i\hat{H}t}|\chi\rangle$ is
the quantum time-evolution of the arbitrary reference state $|\chi\rangle$.
By writing the reference state as a linear combination of the Hamiltonian
eigenstates $|\psi_{j}\rangle$, i.e. $|\chi\rangle=\sum_{j}c_{j}|\psi_{j}\rangle$,
it can be shown that 
\begin{equation}
I(E)=\sum_{j}|c_{j}|^{2}\delta(E-E_{j}).
\end{equation}
Hence, the power spectrum is equal to a sum of delta functions centered
at the vibrational frequencies $E_{j}$. A convenient way to calculate
the formula in Eq. (\ref{eq:Survival}) is given by the time-averaging
semiclassical initial value representation (TA-SCIVR) method.\cite{Kaledin_Miller_Timeaveraging_2003,Kaledin_Miller_TAmolecules_2003,miller1968uniform,miller1971semiclassical,miller1998spiers,miller2001semiclassical,Miller_PNAScomplexsystems_2005,Sun1998,Thoss_Miller_GeneralizedFBIVR_2001,Yamamoto_Miller_Thermalrate_2003}
This is obtained by applying a time-averaging filter to the Heller-Herman-Kluk-Kay
(HHKK) propagator\cite{Heller_FrozenGaussian_1981,Herman_Kluk_SCnonspreading_1984,Miller_HKdemonstration_2002,Kluk_Davis_Highlyexcited_1986,Kay_Multidim_1994,Kay_Integralexpression_1994,Kay_Numerical_1994,Kay_SCcorrections_2006,Grossmann_Xavier_SCderivation_1998,Antipov_Nandini_Mixedqcl_2015,Church_Ananth_Filinov_2017,Church_Ananth_SCinmixedqcl_2019,Buchholz_Ceotto_leakage_2018},
which defines the approximate quantum time evolution. The final TA-SCIVR
expression of Eq. (\ref{eq:Survival}) for a system characterized
by $F$ degrees of freedom reads

\begin{eqnarray}
I(E)=&&
\left(\frac{1}{2\pi}\right)^{F}
\iint d\bm{p}_{0}d\bm{q}_{0}\:\frac{1}{2\pi T}
\nonumber\\
&&
\bigg|\intop_{0}^{T}dt
\,\langle\chi|\bm{p}_{t}\bm{q}_{t}\rangle
e^{i\left(S_{t}(\bm{p}_{0},\bm{q}_{0})+\phi_{t}(\bm{p}_{0},\bm{q}_{0})+Et\right)}
\bigg|^{2},\label{eq:TASCIVR}
\end{eqnarray}
where $T$ is the total simulation time, $S_{t}(\bm{p}_{0},\bm{q}_{0})$
the instantaneous action of the classically evolved trajectory $\left(\bm{p}_{t},\bm{q}_{t}\right)$,
and the phase-space integration is performed on the initial trajectory
momenta $\bm{p}_{0}$ and positions $\bm{q}_{0}$. In the previous
equation $|\bm{p}_{t}\bm{q}_{t}\rangle$ are coherent states with
the following form in position representation\cite{Heller_Cellulardynamics_1991}
\begin{eqnarray}
\langle\bm{x}|\bm{p}_{t}\bm{q}_{t}\rangle=&&
\left(\frac{\det(\boldsymbol{\gamma})}{\pi^{F}}\right)^{\frac{1}{4}}
\nonumber\\
&&
\exp\left[
  -\frac{1}{2}\left(\bm{x}-\bm{q}_{t}\right)^{\mathrm{T}}
  \boldsymbol{\gamma}
  \left(
    \bm{x}-\bm{q}_{t}\right)+i\bm{p}_{t}^{\mathrm{T}}\left(\bm{x}-\bm{q}_{t}
  \right)
\right],
\end{eqnarray}
where $\boldsymbol{\gamma}$ is an $F\times F$ diagonal matrix, whose elements
are chosen to be numerically equal to the harmonic frequencies of
the system. In Eq. (\ref{eq:TASCIVR}), $\phi_{t}(\bm{p}_{0},\bm{q}_{0})$
is the phase of the HHKK prefactor\cite{DiLiberto_Ceotto_Prefactors_2016}

\begin{eqnarray}
&&\phi_{t}(\bm{p}_{0},\bm{q}_{0})=
\nonumber\\
&&
phase\left[
  \sqrt{\frac{1}{2^{F}}
    \left|
      \bm{M_{qq}}+
      \boldsymbol{\gamma}^{-1}\bm{M_{pp}}\boldsymbol{\gamma}-
      i\bm{M_{qp}}\boldsymbol{\gamma}+
      i\boldsymbol{\gamma}^{-1}\bm{M_{pq}}
    \right|}
\right],\label{eq:prefactor}
\end{eqnarray}
where $\bm{M_{ij}}$ with $\bm{i,j}=\bm{p},\bm{q}$ are the elements
of the Jacobian (monodromy) matrix

\begin{align}
\bm{J}=\begin{pmatrix}\bm{M_{pp}} & \bm{M_{pq}}\\
\bm{M_{qp}} & \bm{M_{qq}}
\end{pmatrix}=\begin{pmatrix}\dfrac{\partial\bm{p}_{t}}{\partial\bm{p}_{0}} & \dfrac{\partial\bm{p}_{t}}{\partial\bm{q}_{0}}\\
\\
\dfrac{\partial\bm{q}_{t}}{\partial\bm{p}_{0}} & \dfrac{\partial\bm{q}_{t}}{\partial\bm{q}_{0}}
\end{pmatrix}.\label{eq:jac_matrix}
\end{align}

The determinant $\left|\det(\bm{J})\right|$ is always equal to 1 along the trajectory,
in accordance with Liouville's theorem. The major problem associated
to the TA-SCIVR spectral density calculation is represented by the
computational cost of the Monte Carlo phase-space integration in Eq.
(\ref{eq:TASCIVR}).\cite{Ma_Ceotto_SN2reactions_2018} To overcome
this problem, the multiple coherent states SCIVR (MC SCIVR) has been
introduced. The method relies on the idea that the most important
contribution to the spectrum comes from the trajectories whose energies
are as close as possible to the quantum mechanical eigenvalues.\cite{DeLeon_Heller_SCeigenfunctions_1983}
Thus, in MC SCIVR, the phase-space integral of Eq. (\ref{eq:TASCIVR})
is formally replaced by a sum over the most relevant trajectories,
i.e. those corresponding to the spectral signals of interest. The
MC-SCIVR initial conditions for the $j$-th degree of freedom are\cite{Ceotto_AspuruGuzik_Multiplecoherent_2009,Ceotto_AspuruGuzik_PCCPFirstprinciples_2009}

\begin{equation}
\begin{cases}
q_{0}^{(j)}=q_{eq}^{(j)}\\
p_{0}^{(j)}=\sqrt{(2n_{j}+1)\omega_{j}}
\end{cases},\label{eq:initial_conditions_MCSCIVR}
\end{equation}
being $q_{eq}^{(j)}$ the equilibrium position of the $j$-th mode,
and $\omega_{j}$ its harmonic frequency. Our reference states are
combinations of coherent states of the type 
\begin{equation}
|\chi\rangle=\prod_{j}^{F}\left(|p_{0}^{(j)},q_{0}^{(j)}\rangle+\epsilon_{j}|-p_{0}^{(j)},q_{0}^{(j)}\rangle\right),\label{eq:eps}
\end{equation}
where $\epsilon_{j}=\pm1$ according to which spectroscopic signal
one wants to enhance. For example, a collection of +1 values allows
one to enhance the ZPE signal (together with the even transitions),
while a selected $\epsilon_{j}=-1$ and the remaining $\epsilon_{i\neq j}=+1$
enhances the odd transition of the $j-th$ mode.\cite{Ceotto_AspuruGuzik_Curseofdimensionality_2011}
MC SCIVR has been successfully applied to the study of several systems,\cite{Gabas_Ceotto_Glycine_2017,Ceotto_Tantardini_Copper100_2010,Conte_Ceotto_NH3_2013,Micciarelli_Ceotto_SCwavefunctions2_2019,Micciarelli_Ceotto_SCwavefunctions_2018,Tamascelli_Ceotto_GPU_2014,Buchholz_Ceotto_MixedSC_2016,Buchholz_Ceotto_applicationMixed_2017,Ceotto_Buchholz_SAM_2018,Zhuang_Ceotto_Hessianapprox_2012,Ceotto_Hase_AcceleratedSC_2013,aieta2020anharmonic}
including the different conformers of the glycine amino acid.\cite{Gabas_Ceotto_Glycine_2017}
To improve with respect to the harmonic initial conditions of Eq.s
(\ref{eq:initial_conditions_MCSCIVR}), a preliminary adiabatic switching
warm up can be implemented with the result that frequency estimates
are generally more accurate and complications due to deterministic
chaos are largely avoided.\cite{Conte_Ceotto_ASSCIVR_2019}

However, for very high-dimensional systems, the overlap between the
initial and the evolved wavepacket of\emph{ }Eq. (\ref{eq:Survival})
is smaller and smaller as the dimensionality increases, given that
the reference states are the direct product of monodimensional coherent
states. To overcome this limitation, the Divide-and-Conquer (DC) SCIVR
method has been recently introduced. \cite{ceotto_conte_DCSCIVR_2017,DiLiberto_Ceotto_Jacobiano_2018}
The DC basic idea is to project the system onto lower-dimensional
subspaces. Within these subspaces, it is possible to calculate reduced-dimensionality
spectra. Then, the full-dimensional spectrum can be obtained by convolving
the subdimensional ones. The DC-SCIVR working equation is

\begin{widetext}
\begin{equation}
\tilde{I}(E)=\left(\frac{1}{2\pi}\right)^{D}\iint d\tilde{\bm{p}}_{0}d\tilde{\bm{q}}_{0}\:\frac{1}{2\pi T}\bigg|\intop_{0}^{T}dt\,\langle\tilde{\chi}|\tilde{\bm{p}}_{t}\tilde{\bm{q}}_{t}\rangle e^{i\left(\tilde{S}_{t}(\tilde{\bm{p}}_{0},\tilde{\bm{q}}_{0})+\tilde{\phi}_{t}(\tilde{\bm{p}}_{0},\tilde{\bm{q}}_{0})+Et\right)}\bigg|^{2},\label{eq:DCSCIVR}
\end{equation}
\end{widetext}
where the quantities projected onto a $D$-dimensional subspace have
been indicated with the tilde symbol. The DC-SCIVR approach has been
successfully applied to complex and fluxional systems, like small
water clusters\cite{Ceotto_watercluster_18} and the protonated water
dimer.\cite{Bertaina_Ceotto_Zundel_2019} It is also possible to
implement the multiple coherent states idea into the DC-SCIVR method
by replacing the double integral of Eq. (\ref{eq:DCSCIVR}) with a
sum running on the most relevant trajectories. This method, named
MC-DC SCIVR, can deal with very high-dimensional systems. Notable
applications of MC-DC SCIVR include dipeptide derivatives,\cite{Gabas_Ceotto_SupramolecularGlycines_2018}
nucleobases\cite{Gabas_Ceotto_Nucleobasi_2019} and nucleosides,
\cite{gabas2020semiclassical} and molecules adsorbed on titania
surfaces.\cite{Cazzaniga_Ceotto_skywalker_2020} The most relevant
issue to deal with for a successful DC-SCIVR calculation is the choice
of the optimal subspace decomposition. In fact, all the quantities
appearing in Eq. (\ref{eq:DCSCIVR}) can be exactly projected onto
subspaces, except for the action, because the potential is not in
general separable and its coupling terms significantly change the
action. To project the action, we adopted the following equation 
\begin{eqnarray}
&&\tilde{S}_{t}(\tilde{\bm{p}}_{0},\tilde{\bm{q}}_{0})=\intop_{0}^{t}dt^{\prime}
\nonumber\\
&&
\left[
  \frac{1}{2}\,\tilde{\bm{p}}_{t^{\prime}}^{T}\tilde{\bm{p}}_{t^{\prime}}-
  \left(
    V(\tilde{\bm{q}}_{t^{\prime}},\bm{q}_{t^{\prime}}^{(F-D)})- 
    V(\tilde{\bm{q}}_{eq},\boldsymbol{q}_{t^{\prime}}^{(F-D)})
  \right)
\right],
\end{eqnarray}
which is exact for separable potentials.\cite{ceotto_conte_DCSCIVR_2017}
The projected pre-exponential factor is obtained by substituting the
elements of the $2M\times2M$ sub-block Jacobian matrix of the type
of Eq. (\ref{eq:jac_matrix}) into the prefactor equation (\ref{eq:prefactor}). 

Between the possible criteria that one can adopt to partition the
vibrational space into subspaces, the one that makes each $2D\times2D$
sub-block Jacobian matrix determinant closer to 1 is the less severe
approximation for the reduced dimensionality spectra. We have called
this procedure the ``Jacobi method''.\cite{DiLiberto_Ceotto_Jacobiano_2018}
By recalling Liouville's theorem, this vibrational space subdivision
is the one that minimizes the energy exchange between subspaces, because
an ideal partition where each sub-block determinant is equal to unity
preserves the energy within each subspace. More specifically, we compute
the Jacobian matrix (Eq. (\ref{eq:jac_matrix})), which, in turn, is
computed during the dynamics according to the prescription given by
Brewer et al. \cite{Brewer_Manolopoulos_15dof_1997}.

The main goal of this work is to find an efficient method for the
subdivision of the Jacobian matrix of Eq. (\ref{eq:jac_matrix}) into
a number $n_s$ of subdimensional Jacobian matrices, each containing
a cluster of normal modes, such that each subspace evolution $\left(\tilde{\bm{p}}_{t},\tilde{\bm{q}}_{t}\right)$
is the closest possible to satisfy Liouville's theorem. In other words,
an optimal choice of the normal mode clustering would lead to $\prod_{i}^{n_s}\det\mathbf{\tilde{J}}_{i}$
as close as possible to 1, where $\tilde{\mathbf{J}}_{i}$ is the
Jacobian matrix of subspace $i$, and it is extracted from the full-dimensional
Jacobian simply selecting all the entries involving the modes that
belong to subspace $i$. In our procedure, we compute the Jacobian
matrix at every time step along a test trajectory, which starts from
the equilibrium position of the atom coordinates and with initial
kinetic energy equal to the harmonic zero point energy. Originally,\cite{DiLiberto_Ceotto_Jacobiano_2018}
we presented a hierarchical search of the most frequently selected
subspaces along the test trajectory in a two-step procedure. First,
for each possible dimensionality $1\dots D<F$, all the ${F \choose D}$
possible subspaces were considered and the most frequently chosen
along the dynamics were saved. Then, the absolute value of the deviation
from 1 of each subspace Jacobian was computed at each time step, and
the optimal subspace for each dimensionality was declared the one
with the smallest average deviation. The subspace associated with
the overall smallest deviation was selected and the whole procedure
reiterated on the remaining degrees of freedom until all modes had
been included into a subspace. This approach has still a non convenient
computational scaling cost which is proportional to $\sum_{D=1}^{F-1}D\binom{F}{D}$,
since all subspace combinations need to be tested. Furthermore, it
is hierarchical and thus prone to find local optima for the global
subdivision. In this study, we develop and test an algorithm able
to find the global optimal subspaces with lower computational efforts
and given a constrained maximum subspace dimensionality $\bar{D}$.
This maximum dimensionality constraint can be freely chosen at the
beginning of our proposed procedure. It is useful because, as anticipated,
it comes from the necessity to perform semiclassical calculations
below a certain dimensionality to get sensible results.

\subsection{Probability Graph-Evolutionary Algorithms (PG-EA)}

\label{sec:GA}

Here we introduce a combined Probability Graph and Evolutionary Algorithm
(PG-EA) approach to find the best vibrational space subdivision according
to Liouville's criterion explained above. Evolutionary algorithms
emulate the natural selection of an initial population, where the
``fittest'' individual is the most likely to survive and its genes
to be inherited by the next generation. In GAs jargon, the population
is composed of chromosomes, which are collections of fitness parameters,
each one called a gene. There is no obvious or required way to represent
the genes and there are many valid choices. At each epoch, all chromosomes
are evaluated and sorted according to their fitness score. First,
a fraction of the best individuals give birth to a set of newborn
chromosomes by mixing and mutating genes during the crossover and
mutation processes. Then, the new chromosomes take the place of those
individuals that are least fit to survive, so that the next epoch
would be enriched by the more fitted chromosomes.

In our case, each chromosome represents a possible clustering of vibrational
degrees of freedom into subspaces to compose the full vibrational
space. The collection of all the chromosomes provides many 
possible subdivisions of the full-dimensional vibrational space. Each
chromosome is evaluated by an appropriate score function that rates
the individual's fitness. The fitness function is evaluated after
time evolution of the Jacobian matrix along a test trajectory, which
in our case is the trajectory that evolves from the equilibrium geometry
with the energy of the vibrational ground state. Given the consideration at the
end of Sec. \ref{subsec:Semiclassical-spectra}, we propose the following
fitness function for a possible collection $C$ of normal modes subdivisions

\begin{equation}
f(C)=
\dfrac{1}{N}
\sum_{\mathrm{steps}}^{N}
\sum_{s\in C}
\left|
  1-\left|\det(\tilde{\bm{J}}_{s})\right|
  \right|,\label{eq:GAfitness2}
\end{equation}
where the external sum is over the $N$ time steps.
We prefer
Eq. (\ref{eq:GAfitness2}) with respect to a possible fitness function,
such as $|1-\prod_{i}^{n_s}\det\mathbf{\tilde{J}}_{i}|$ as employed
in Ref.\citenum{DiLiberto_Ceotto_Jacobiano_2018},
because in the latter case there could be
a compensation of error that the internal sum in Eq. (\ref{eq:GAfitness2})
avoids. More specifically, the optimal criterion in Eq. (\ref{eq:GAfitness2})
is satisfied when all the subspaces have the determinant of the Jacobian
closest to +1 in modulus, at every trajectory step. Furthermore, this
function rewards preferentially a chromosome made of few large subspaces
over one made of several small subspaces because any new term in the
internal sum over C is addictive and positive. We prefer to have large
subspaces to account for as many normal modes couplings as possible.

Once an initial guess of possible chromosomes is given, we need a
probability distribution function to generate the new chromosomes,
i.e. the new mode subdivision into subspaces. We propose our own customized
Evolutionary Algorithm inspired to GAs for updating the probability
  distribution $\bm{\Phi}(\tau)$ from which newborn chromosomes are
sampled at a certain epoch $\tau$. 

\begin{figure}[H]
\centering
\includegraphics[width=1.0\columnwidth]{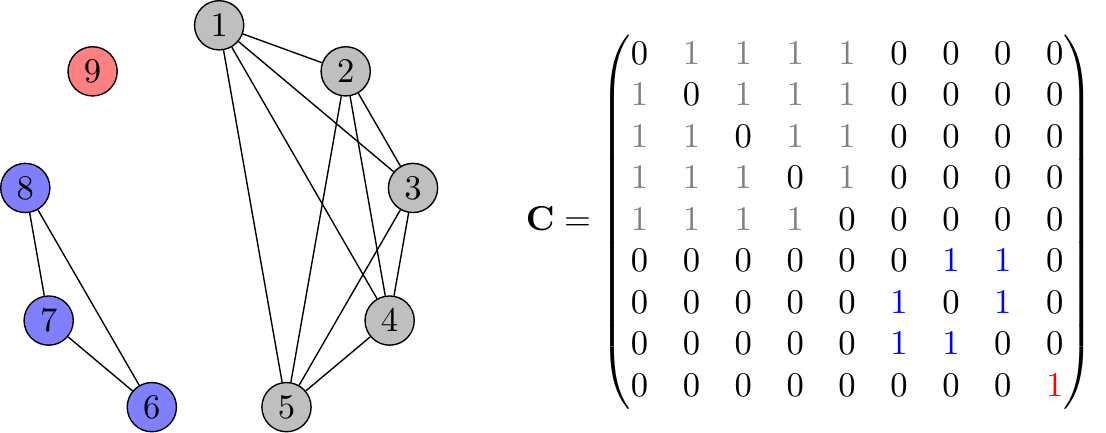}
\caption{Chromosome expressed in a cluster graph representation. Here, 9 normal
modes, corresponding to the graph vertexes, are grouped into three
subspaces (gray, blue, and red), containing 5, 3, and 1 modes, respectively.
Each subspace is a fully connected graph (a clique), and there is
no interaction between any two subspaces. On the right, the corresponding
adjacency matrix is colored accordingly.}
\label{fig:cluster_graph} 
\end{figure}

First, we represent a chromosome
as the adjacency matrix of an unweighted cluster graph, that is a
graph where each connected component is a clique (i.e. a fully connected
subgraph), as reported in Figure \ref{fig:cluster_graph}. The vertexes
are the normal modes that are connected only if they fall into the
same subspace. This representation minimizes the redundancy of information,
since cliques are invariant to vertex permutation and a cluster graph
is invariant to clique permutations.

All the information about the subspaces is codified in the adjacency
matrix $\bm{C}$. It is defined by $C_{ij}=1$ only if modes $i$
and $j$ are in the same subspace, otherwise $C_{ij}=0$. $C_{ii}=1$
only if mode $i$ is in a one-dimensional subspace. In the end, we
have increased the problem variables from the $F$-dimensional redundant
representation (such as in the linear representation (1,2,3,4,5)(6,7,8)(9)=(2,4,1,5,3)(8,7,6)(9)=...)
to the $F(F+1)/2$-dimensional non-redundant one. The $F$-dimensional
representation is redundant because any mode permutation within a
given subspace leaves the subspace unaffected. Our adjacency matrix
representation is invariant to row and column permutation within the
subspace block. Hence, every subspace configuration has a unique adjacency
matrix representation. Furthermore, our adjacency matrix is symmetric,
thus completely defined by its F(F+1)/2 lower (upper) triangular elements.

Secondly, we customize the crossover and mutation operators to mix the
chromosomes with simple arithmetic rules and store the genetic information
in a matrix of weights $\bm{\Phi}(\tau)$, which represents the probability
distribution of the mixed chromosomes at the evolution epoch $\tau$.
We define the crossover $\mathcal{X}$ of a couple of chromosomes
$C$ and $C^{\prime}$ as a weighted average of their adjacency matrices
\begin{equation}
\mathcal{X}=\dfrac{1}{w_{C}+w_{C^{\prime}}}\left(w_{C}\bm{C}+w_{C^{\prime}}\bm{C}^{\prime})\right),\label{eq:crossover}
\end{equation}
where $\mathbf{C}$ and $\bm{C}^{\prime}$ are adjacency matrices
and the weights $w_{C}$ and $w_{C^{\prime}}$ depend on the chromosome
fitnesses. The pure mutation $\mathcal{M}$ of a chromosome $C$ with
probability $\mu$ is 
\begin{equation}
\mathcal{M}=\dfrac{1}{1+F\mu}\left(\mathbf{1}\mu+\bm{C}\right),\label{eq:mutation}
\end{equation}
where $\mathbf{1}$ is the square matrix of ones and $F$ is the number
of genes. In our approach each vibrational normal mode corresponds
to a gene, as illustrated in the previous example of Fig. 
\ref{fig:cluster_graph}.
Both crossover and mutation have the basic property of scattering
the gene probability. More specifically, the crossover distributes
the probability among the genes expressed in $C$ and $C^{\prime}$,
depending on their fitness, and the mutation distributes it among
every possible outcome, independently of the fitness and the expression.
The combination of the crossover and mutation processes is obtained
by the subsequent application of the two operations: the mutation
can be equivalently applied to the single chromosomes previous to
undergoing crossover, or to the result of the crossover process. In
addition, considering that we use $m$ chromosomes for the optimization,
only an elite fraction $\eta<1$ of these are the fittest chromosomes
that undergo the crossover and mutation processes. Eventually, the
resulting probability distribution at epoch $\tau$, $\bm{\Phi}(\tau)$,
is: 
\begin{equation}
\mathbf{\Phi}(\tau)=\dfrac{1}{1+F\mu}\left[\mathbf{1\mu}+\dfrac{1}{\sum_{i}^{m\eta}w_{i}}\sum_{i=1}^{m\eta}w_{i}\bm{C}_{i}\right],\label{eq:prob_distr}
\end{equation}
where $\text{C}_{i}$ is the $i$-th chromosome adjacency matrix and
the new probability distribution is essentially generated by suitably
mixing the adjacency matrices of the previous generation. This is
the machine learning part, where the algorithm, epoch by epoch, learns
the optimal probability distribution from an evolving population of
chromosomes. For this work we use the simple weighting scheme $w_{i}=(m\eta-i)/m\eta$
with the resulting normalization constant $\sum_{i}^{m\eta}w_{i}=(m\eta-1)/2$,
the elite fraction is $\text{\ensuremath{\eta}}=0.4$, while the mutation
probability $\mu$ and the number of chromosomes will be specified
below case by case. $\bm{\Phi}(\tau)$ is updated at every epoch and
contains the average genetic material of the previous generation of
chromosomes according to Eq. (\ref{eq:prob_distr}).

To sample new chromosomes from the probability distribution, $\mathbf{\Phi(\tau)}$
must be normalized. For the expression in Eq. (\ref{eq:prob_distr}),
this means that $\mathbf{\Phi(\tau)}$ has to be symmetric and doubly
stochastic, i.e. with rows and columns summing up to 1, so that we
can consider $\mathbf{\Phi(\tau)}$ a weighted undirected graph to
sample from. To enforce the doubly stochastic property and, at the
same time, retain the symmetry, we rely on Sinkhorn's theorem,\cite{sinkhorn1967}
which ensures that there exist two diagonal matrices $\mathbf{R}$
and $\mathbf{S}$ such that $\mathbf{R}\bm{\Phi}(\tau)\mathbf{S}$ is doubly
stochastic. $R_{ii}$ and $S_{jj}$ are found by repeatedly and alternatively
normalizing the rows and the columns of $\bm{\Phi}(\tau)$, according to
the updates
\begin{equation}\label{eq:sinkhorn}
\begin{split}
R_{ii} & =\dfrac{1}{\sum_{j}\Phi_{ij}(\tau)S_{jj}}\quad\ \forall\ i\\
S_{jj} & =\dfrac{1}{\sum_{i}\Phi_{ij}(\tau)R_{ii}}\quad\ \forall\ j\ ,
\end{split}
\end{equation}
with $S_{jj}$ initialized to 1 for all $j$. $\mathbf{R}$ and $\mathbf{S}$
will converge, up to a small threshold $\varepsilon$, after an unspecified
number of iterations\cite{sinkhorn1967}. In all the applications
described below we use $\varepsilon=10^{-8}$ on each element of $\mathbf{R}$
as and $\mathbf{S}$, which is always satisfied in less than 100 iterations.

Third, we need to elaborate a procedure for obtaining the newborn
chromosomes from the symmetric and doubly stochastic probability matrix
$\mathbf{\Phi}(\tau)$. To sample representative cluster graphs from
$\bm{\Phi}(\tau)$, we propose a sampling procedure to generate a
population which reflects the original distribution:

\label{list:sampling} 
\begin{enumerate}
\item generate the random numbers $r_i, i=1,2,\dots F$ and sample independently
the chances of each mode to be in a subspace alone. If 
$r_{i}<\Phi_{ii}(\tau)$
then the normal mode $i$ is in a subspace alone;
\item iterate on the leftover modes in a random order: if the $k$-th
mode is already joined with another mode, then continue with the next
one, otherwise sample the edge between modes $k$ and $j\neq k$ with
a random number and join them with a probability given by the matrix
element $\Phi_{jk}(\tau)$. If the $k$-th cannot be joined to any $j$
(for instance, because each of them is in a one-dimensional subspace)
it stays in a subspace by itself;
\item identify the connected components of the sampled graph and complete
them, obtaining the cluster graph for a newborn chromosome.
\end{enumerate}
Note that before step 3, the procedure samples tree graphs, which
means that there are no redundant sampling steps. Each chromosome
sampled with this procedure may be weakly biased anyway and the random
shuffle of the mode order in step 2 is required to make the sampled
population representative and the overall sampling unbiased. To achieve
step $3$ we look for a basis of the Laplacian matrix Kernel, with
the Laplacian matrix defined as $L_{ij}=-C_{ij}+\delta_{ij}\sum_{j}C_{ij}$.
Since $\sum_{j}L_{ij}=0$ by definition, the vector of ones always
belongs to $Ker(\mathbf{L})$, i.e. to the collection of vectors $\text{\textbf{x}}$
such that $\mathbf{Lx}=\boldsymbol{0}$. Furthermore, if the graph
is disconnected, $\bm{L}$ can be rearranged to be block
diagonal by swapping row and column indexes, with each block being
the Laplacian of the corresponding connected subgraph. Hence each
basis vector of $Ker(\bm{L})$ is 1 on the entries of the
connected vertices and 0 elsewhere. The sum of all basis vectors is
the vector of ones. To practically find a basis for $Ker(\mathbf{L})$
we solve the linear equation $\mathbf{Lx}=\boldsymbol{0}$ by applying
Gaussian Elimination\cite{higham2011gaussian} to the augmented matrix
$\mathbf{L}|\mathbf{I}$, with output $\mathbf{L_{rref}|B}$, where
$\mathbf{L_{rref}}$ is the reduced row echelon form of $\mathbf{L}$.
The rows $\mathbf{x}$ in $\mathbf{B}$ corresponding to the row indices
where $\mathbf{L_{rref}}=\boldsymbol{0}$ do solve the linear equation
and hence form a basis for the kernel.

To measure the likelihood of a subspace $s$ of size $D$ sampled
from $\bm{\Phi}(\tau)$, we sum the edge products of all the possible trees
that span the clique (subspace), as 
\begin{equation}
p(s,\tau)=
\frac{(D-1)^{D-1}}{D^{D-2}}
\sum_{\bm{T}\in span(\tilde{\bm{\Phi}}_{s}(\tau))}
\prod_{e=1}^{D}
T_{e},\label{eq:likelihood_cay}
\end{equation}
where $T_{e}$ is the edge of the tree graph $\bm{T}$, which spans
the subspace probability distribution $\tilde{\bm{\Phi}}_{s}(\tau)$ (that
is the probability distribution considering only the modes in $s$).
The first factor is a normalization constant so that $p(s,\tau)$
does not depend on the subspace size and it is maximized to 1 when
$\tilde{\bm{\Phi}}_{s}(\tau)$ is uniform. $D^{D-2}$ is the number of spanning
trees for a clique according to Cayley's formula.\cite{cayley1889theorem}
$p(s,\tau)$ measures the degree of convergence towards the chosen
subspace $s$, such that, as $p(s,\tau)$ approaches unity, the population
becomes more and more uniform and eventually the algorithm stops learning.
The brute force application of Eq. (\ref{eq:likelihood_cay}) is
out of reach for large subspaces ($D>\approx10$), therefore, we use
it only to check the algorithm progression towards an optimal solution
of the small systems described below. Furthermore, GAs in general,
and PG-EA in particular, do not require a full convergence of the
population for the solution to be satisfactory. On the contrary, if
the solution is unknown or hard to find, a homogeneous population
is undesirable, as it kills diversity and damps the optimization.

\begin{figure*}[t]
\centering
\includegraphics[width=1.0\textwidth]{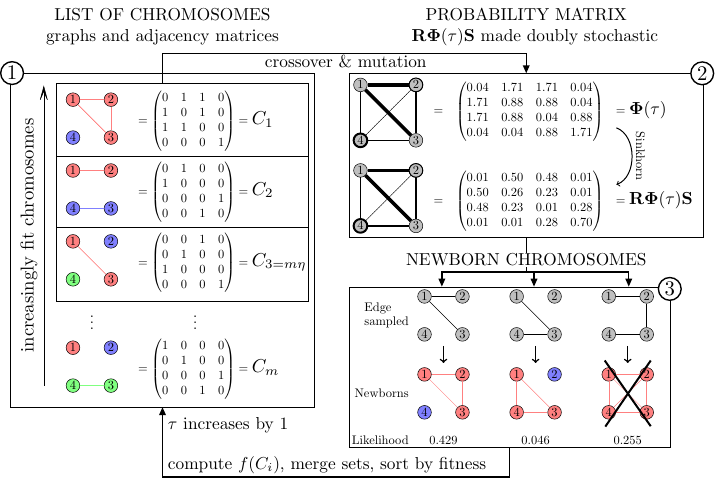} 
\caption{\label{fig:PGEA-pic}A numerical example of the PG-EA for a four dimensional
system. The procedure is broken down into 3 steps. The inner rectangle
in panel 1 includes the $m\eta$ elite chromosomes which will be employed
for mutation and crossover in panel 2.}
\end{figure*}

In Fig. (\ref{fig:PGEA-pic}) we report a four normal mode example
to show how the PG-EA algorithm works in practice. First, as we do
in all our simulations, the initial probability distribution is set
as $\Phi_{ij}(0)=1/F\ \forall\ i,j$. Then, we generate the initial chromosome
population according to the sampling procedure described at points
1-3 above. Each chromosome graph is reported together with the corresponding
adjacency matrix in panel 1 in the left side of Fig. (\ref{fig:PGEA-pic}).
The chromosomes $C_{j}$ are ordered by the fitness score,
which is calculated using the function $f\left(C\right)$ 
(pay attention that according to our definition of the fitness function,
Eq. \ref{eq:GAfitness2}, the preferred chromosomes are those with lower 
fitness score).
The chromosome fraction $m\eta$ undergoes crossings and mutations.
We reject (i.e. apply an infinite penalty) to any subspace which has
a dimension larger than the largest subspace value $\bar{D}$ that one fixes
\emph{a priori}. Since assigning the fitness score is the most expensive
step, it is advisable to build a score database, \emph{i.e.} a list
of already known chromosomes with their fitness value, so that, whenever
a known chromosome is encountered its score does not have to be recomputed.
After undertaking crossovers and mutations according to Eq. (\ref{eq:prob_distr}),
a new (non-normalized) probability distribution is generated (panel
2 on the right side of Fig. \ref{fig:PGEA-pic}). After transforming
the new probability distribution into a symmetric and doubly stochastic
distribution matrix using Sinkhorn's algorithm, we generate the
newborn chromosomes according to the sampling procedure of points
1-3 described above. A likelihood coefficient is calculated according
to Eq. (\ref{eq:likelihood_cay}). In the lower right part of Fig.
\ref{fig:PGEA-pic}, the 4-dimensional solution is rejected because
its dimensionality is greater than the largest subspace value $\bar{D}$,
which in this numerical example has been fixed to be $\bar{D}=3$. Finally,
a fitness coefficient is attributed to each chromosome and we are
back to panel 1 for the next iteration. At each epoch, the fittest
$i$-th chromosome, i.e. the one with the lowest $f\left(C_{i}\right)$
value, provides the graph with the so far optimal normal modes arrangement
into subspaces.

\subsection{2-mode interaction method}

\label{sec:JG} As an alternative, we propose an approximate and computationally
cheaper method to deal with large molecules when the computational
cost of PG-GA is prohibitive or in instances in which one can reasonably
assume that for each normal mode the coupling is mainly due to the
interaction with just a second mode.

In this alternative approach, we first compute a 2-mode coupling network,
where each vertex $i,j$ is a normal mode and each edge is weighted
by the two-mode Jacobian determinant $G_{ij}=\det(\tilde{\bm{J}}_{ij})$.
$\tilde{\bm{J}}_{ij}$ is a $4\times4$ matrix containing all the
partial derivatives between the phase-space momenta and positions
of modes $i$ and $j$ with respect to the initial conditions. For
each full-dimensional Jacobian matrix $\bm{J}$ along the trajectory,
we evaluate the determinant of every two-mode combination, $G_{ij}$.
Then, we compute the distance matrix $\bm{E}=|\bm{G}-\bm{1}|$,
which measures how large is the error done by assuming that the relevant
interaction is only between the couple (i,j) of modes while other
interactions are disregarded. Specifically, when $E_{ij}=0$, then
modes $i$ and $j$ are fully correlated and uncoupled to any other
mode. $\bm{E}$ is computed at every time step of the test trajectory
and all $\bm{E}$ matrices are averaged into a single distance matrix representative
of the whole trajectory.

Then, we employ an agglomerative hierarchical clustering technique
called Weighted Pair Group Method with Arithmetic mean (WPGMA) \cite{sokal1958statistical}
to cluster the normal modes using the information encoded in $\bm{E}.$
The algorithm produces a dendrogram where each branching is an optimal
subspace. Among the several hierarchical clustering techniques available
in the literature, we choose WPGMA because it provides results which
are the closest to the exact ones for the model systems considered below.
WPGMA clustering is iterative and hierarchical. To start, each mode
is in a subspace by itself. Then, at every iteration, the two ''closest''
subspaces are merged into one and the dendrogram profile shows a link.
The distance between the newly formed subspace $(j\cup k)$ and a
given subspace $i$ is calculated as the arithmetic mean of the distances
from the newly merged subspaces $j$ and $k$: 
\begin{equation}
E'_{i(j\cup k)}=\dfrac{E_{ij}+E_{ik}}{2}.
\end{equation}
The procedure goes on until a maximum distance criterion has been
met, i.e. until all modes fall into one large subspace.

The whole process is represented by a dendrogram, where each node
corresponds to a subspace and each edge represents the link between
two subspaces. At the root of the dendrogram there is the full-dimensional
system, which contains all modes. The leaves are the subspaces containing
one mode only. The distance $E_{ij}$ of every update is a measure
of how close the linked subspaces are. Finally, this process generates
a number of arrangements of normal modes at different level of the
tree, and for every such arrangement we measure the fitness score,
i.e. the full-dimensional Jacobian factorization error with Eq.
(\ref{eq:GAfitness2}), along
with $E_{ij}$.

\section{Results\label{sec:Results}}

This section presents our results and it can be divided into three
parts. In Sec. \ref{sec:model_systems}, we show how PG-EA and the
2-mode interaction method are effective when applied to model systems
like coupled Morse oscillators with non-trivial coupling topologies
but obvious mode separations. In subsection \ref{sec:molecules},
we show that we can improve spectra accuracy with respect to previous
calculations where the hierarchical subspace optimization originally
proposed was adopted.\cite{DiLiberto_Ceotto_Jacobiano_2018} Finally,
in subsection \ref{sec:NMA} we show that PG-EA allows us to apply
the DC-SCIVR method and select the subspaces with the Jacobi criterion
even for simulation of mid-large molecules such as the 12-atom \emph{trans}-N-Methylacetamide.
Remarkably, it would not have been possible to accomplish this task
with a brute force combinatorial approach.

\subsection{Model systems}

\label{sec:model_systems} 
\begin{figure}[h]
\centering
\includegraphics[width=1.0\columnwidth]{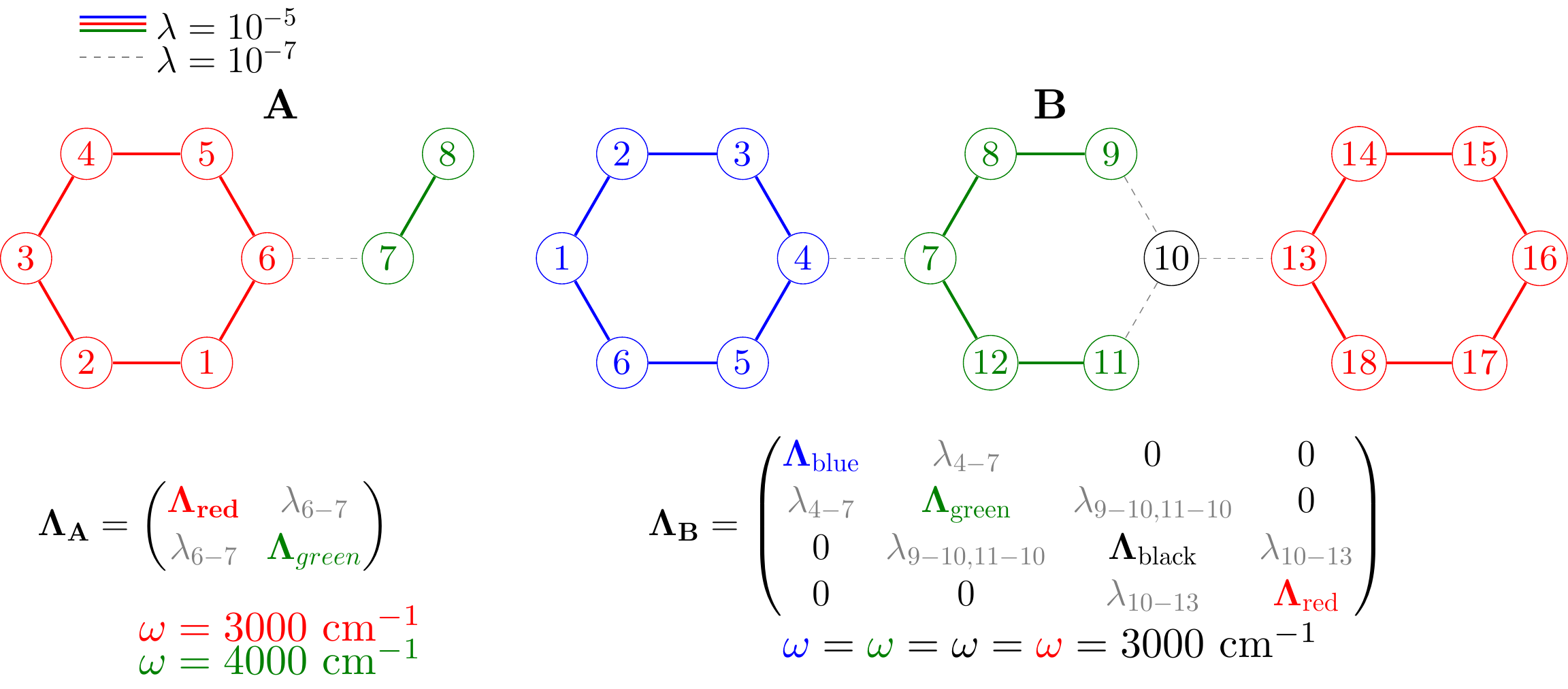}
\caption{Toy model systems with different coupling topologies. The circled
numbers represent Morse oscillators and the edges their couplings.
When two oscillators are not connected by an edge, the relative coupling
matrix element $\lambda_{ij}$ is zero. The oscillator frequencies
$\omega$, reported below each coupling matrix, are 3000 $\cmr$ except
for modes 7 and 8 of topology \textbf{A}, for which $\omega=4000\cmr$.
Note that for topology \textbf{A} reported on the left, oscillators
1 and 5 are equivalent and they produce the same signal, as well as
2 and 4. Similar symmetry considerations can be applied to topology
\textbf{B}.}
\label{fig:toy-models}
\end{figure}

To preliminarily test our algorithms, we consider the arrangements of
$F$ coupled Morse oscillators \textbf{A} and \textbf{B} in figure
\ref{fig:toy-models}. Each oscillator experiences the following Morse-type
potential 
\begin{eqnarray}
V^{morse}=&&
  \sum_{i=1}^{F}D_{e}
  \left(
    1-e^{-\omega_{i}(2D_{e})^{-1/2}(q_{i}-q_{eq,i})}
  \right)^{2}+\nonumber
  \\
&&\sum_{i=1}^{F-1}
  \sum_{j>i}^{F}
  \lambda_{ij}(q_{i}-q_{eq,i})(q_{j}-q_{eq,j}),
  \label{eq:morse_pot}
\end{eqnarray}
where the dissociation energy 
$D_{e}=38293 \cmr$
and the
equilibrium position $q_{eq}=1.4\:\text{a.u.}$ are valid for all
$F$ degrees of freedom. According to the coupling graphs and matrices
$\mathbf{\Lambda}$ schematically represented in Figure \ref{fig:toy-models},
the oscillators might be uncoupled (no edge), weakly coupled 
($\lambda=10^{-7}\:\text{a.u.}$,
dashed edge) or strongly coupled ($\lambda=10^{-5}\:\text{a.u.}$,
solid edge). We devise two topologies (\textbf{A} and \textbf{B})
to provide non-trivial examples. \textbf{A} has oscillator frequency
$\omega=3000\:\text{cm}^{-1}$ for the oscillators 1 through 6 and
$\omega=4000\:\text{cm}^{-1}$ for oscillators 7 and 8; \textbf{B}
has all oscillators with the same frequency $\omega=3000\:\text{cm}^{-1}$.
In both cases the correct separation into subspaces is unique.

\begin{figure*}[t]
\begin{centering}
\includegraphics[width=0.9\textwidth]{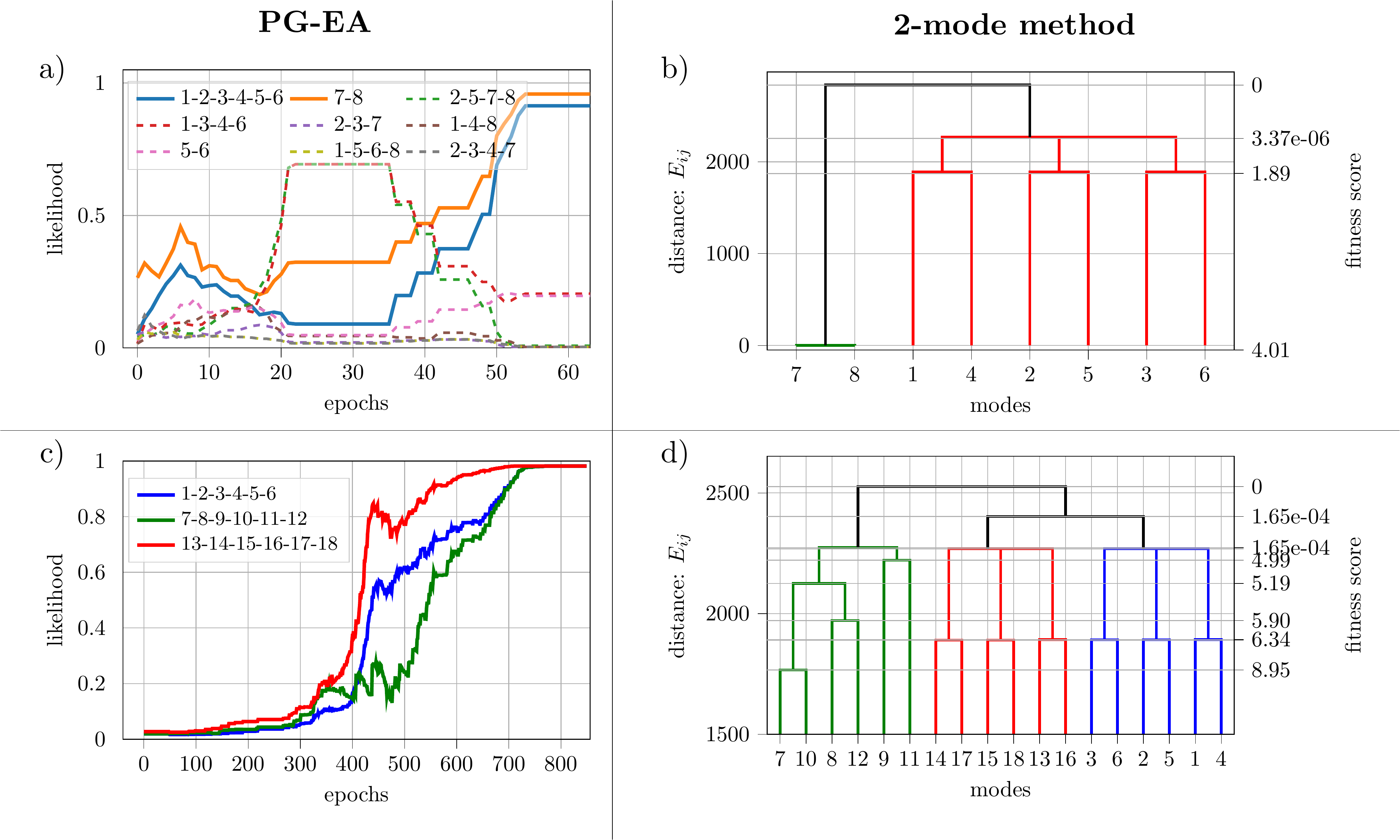}
\par\end{centering}
\caption{Vibrational modes subdivision optimizations of the model systems in
Fig. (\ref{fig:toy-models}) using PG-EA (left panels (a) and (c))
and the 2-modes interaction method (right panels (b) and (d)). In
the top panels ((a) and (b)) the subspace optimizations of topology
\textbf{A} is effectively achieved using both methods. The dendrogram
(panel (b)) is colored to highlight the least error branching. In
the bottom panels (panels (c) and (d)) the subspace optimization is
effectively reached by both methods for topology \textbf{B}.}
\label{fig:toy-models-opt} 
\end{figure*}

Both PG-EA and the 2-mode interaction method separate system \textbf{A}
correctly. In PG-EA we use $m=50$ chromosomes, mutation probability
$\mu=0.001$, crossover fraction $\eta=0.4$ and 70 epochs,
with the constraint that the maximum dimension is $\bar{D}=6$. The likelihood
of the optimal subspaces calculated using Eq. (\ref{eq:likelihood_cay})
is plotted against the epochs in panel (a) at the top left of Fig.
\ref{fig:toy-models-opt}, showing that the population quickly converges
to the unique global optimum represented by the continuous lines.
The 2-mode interaction method provides three choices for the subspaces,
the best of which is the global optimum (1,2,3,4,5,6)(7,8), reported
in the first branching of the dendrogram in panel (b) at the top right
of Fig. \ref{fig:toy-models-opt}. This global optimum has a Jacobian
factorization error of about $3.37\cdot10^{-6}$. The example of topology
\textbf{B} is much more challenging, as it is an 18-dimensional system
divided in 4 loosely connected regions, one of which containing a single
mode. As expected, it turns out that the best subspace division has oscillator
10 (black subspace in Fig. \ref{fig:toy-models}) joined together
with five oscillators (7-12, green subspace), since there is one term
less in the fitness function summation with respect to the case in which 
oscillator 10 is left isolated. In panel (c) at the bottom
left of Fig. (\ref{fig:toy-models-opt}), PG-EA provides the optimal
desired solution using $m=300$ chromosomes, $\mu=0.1$, $\eta=0.4$
and 1000 epochs, with the constraint that
the largest subspace dimension is $\bar{D}=10$. In panel (d) at the bottom
right of Fig. \ref{fig:toy-models-opt}, the 2-mode interaction method
also provides the optimal solution, as shown in the upper part of
the dendrogram with the smallest Jacobian factorization error.

\begin{figure}[h]
\centering
\includegraphics[width=1\columnwidth]{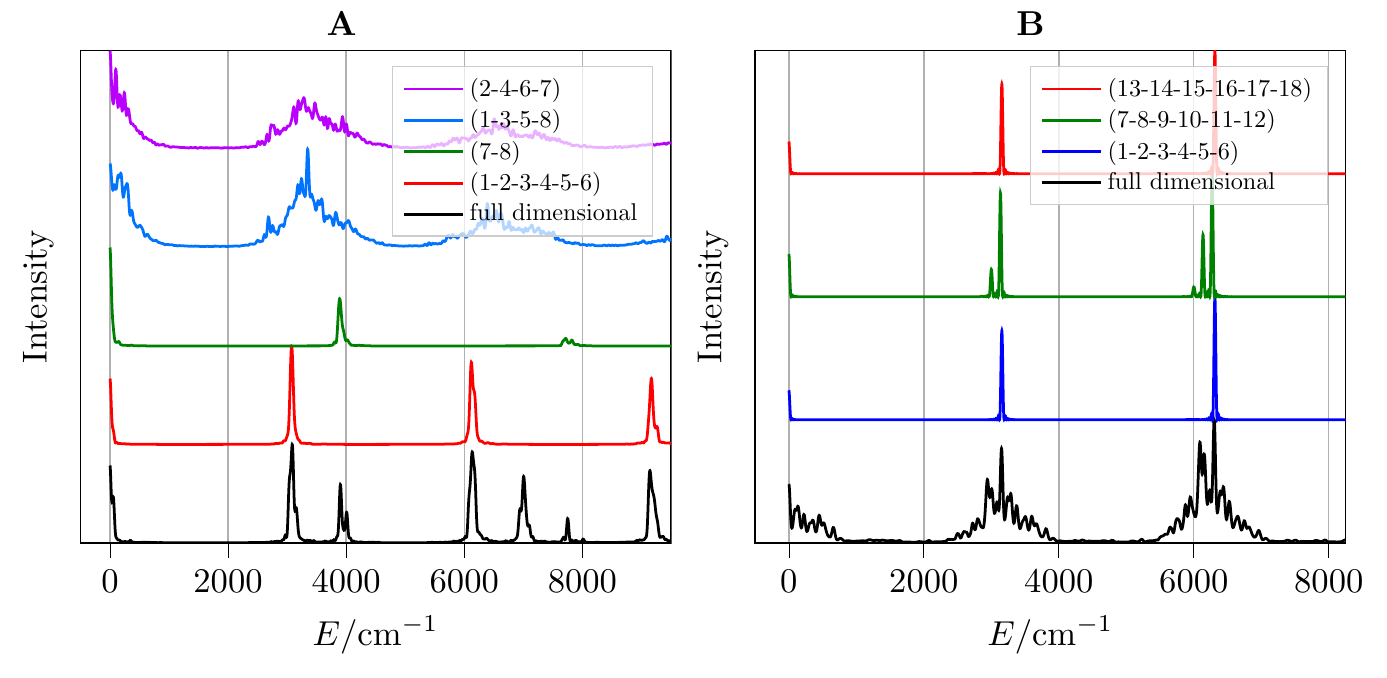}
\caption{Spectra of the coupling topologies \textbf{A} and \textbf{B} of Morse
oscillators with ZPE signals shifted to 0 $\mathrm{cm^{-1}}$. Left
panel for system \textbf{A}: in black the full-dimensional spectrum;
in red and green the best subspaces (ring and segment respectively
with reference to figure \ref{fig:toy-models}), while the blue and
purple are the two worst subspaces. Right panel for system \textbf{B}:
in blue, dark green and red the best subspaces; in black the full-dimensional
calculation.}
\label{fig:spectra_toy-models} 
\end{figure}

Now, one may wonder if these subdivisions are indeed the most suitable
ones for DC-SCIVR spectroscopic calculations. In Figure \ref{fig:spectra_toy-models}
we show that DC SCIVR can account properly for most of the spectral
features of these systems, if the subspaces are chosen accordingly
to the algorithms described above. For example when choosing the subspace
separation (1,3,5,8)(2,4,6,7), which is the least fit for case \textbf{A}
i.e. it has the largest fitness score in case of requiring 2 subspaces
only, the corresponding spectra are quite noisy as shown in Figure
\ref{fig:spectra_toy-models}. Furthermore, phantom signals are observed,
for example at $2686\:\text{cm}^{-1}$. Conversely, the spectra of
the subspaces suggested by both our algorithms, which are reported
with green and red lines, are without noise to the naked eye. However,
we notice that the signal originated from a combination band of modes
from different subspaces at $7006\:cm^{-1}$ is too weak in this scale
to be observed. This is not a drawback of the algorithms proposed
in this work, but a known feature of the DC-SCIVR method in predicting
mixed overtones originated from modes belonging to different subspaces.

On the right part of Fig. \ref{fig:spectra_toy-models} we show the
optimal subspace spectra for system \textbf{B}. In this case, the
system is too large to have a well-converged semiclassical full-dimensional
TA-SCIVR spectrum as shown by the black continuous line spectrum\cite{Ceotto_AspuruGuzik_Curseofdimensionality_2011}.
Instead, it is possible to recover the most significant spectroscopic
features of the system with a DC-SCIVR calculation based on the optimal
subspaces suggested by the algorithms.

\subsection{The \ce{CH4} Molecule}

\label{sec:molecules} This section further confirms the ability of
the proposed algorithms to find optimal subspace separations for DC-SCIVR
calculations when applied to real systems. We show that our techniques
can reproduce and improve the DC-SCIVR spectra even for small molecules.
We consider \ce{CH4} as the case system. Methane vibrational spectrum
is a tough challenge for DC SCIVR because the molecule is characterized
by highly chaotic dynamics and high symmetry which is difficult to
recover if a proper subspace partition is not implemented. We simulate
180000 trajectories 30000 a.u. long and each trajectory is rejected
if during the dynamics $||\det(\mathbf{J}^{T}\mathbf{J})|-1|>10^{-5}$.
The initial trajectory conditions are sampled from the Husimi distribution
centered in phase space at $(\sqrt{\boldsymbol{\omega}},\bm{q}_{eq})$,
while gradients and Hessian matrices are computed by finite differences
with infinitesimal displacements equal to $10^{-3}$ a.u. for all
modes.

\begin{figure*}[t]
\centering
\includegraphics[width=1\textwidth]{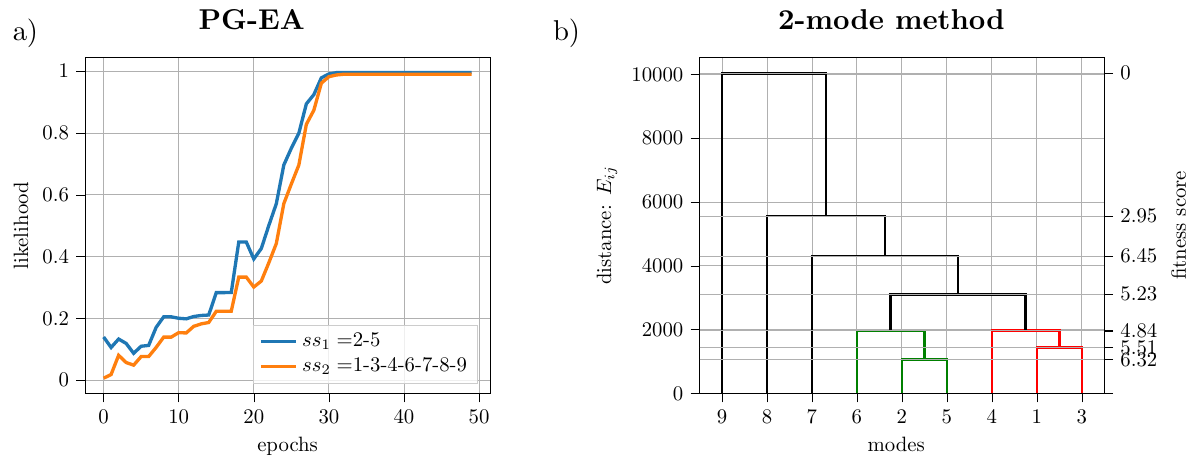}
\caption{Likelihood of the optimal subspaces of methane during PG-EA optimization
(panel (a)) and dendrogram for separation with 2-mode approximation
method (panel (b)). Note that the 2-mode approximation leads to a
very different result.}
\label{fig:dendrogram_methane}
\end{figure*}

We use the Force-Field by Lee, Martin and Taylor \cite{lee_taylor_PESch4_1995}
which takes into account the symmetry relations of cubic and quartic
force constants. \cite{gray1979anharmonic,raynes1987calculations}
The same PES and the hierarchical subspace optimization with the Jacobi
method was employed in a previous work of the group.\cite{ceotto_conte_DCSCIVR_2017}
For this system PG-EA successfully converges with the constraint $\bar{D}\le7$,
which leads to the optimal couple of subspaces (2,5)(1,3,4,6,7,8,9),
with a fitness score of about 0.71. The three subspaces (1)(2,3)(4,5,6,7,8,9)
were selected in a previous work of the group \cite{ceotto_conte_DCSCIVR_2017}
using the Jacobi criterion but looking for optimal subspaces with
a brute force hierarchical approach and constraining the largest subspace
to be six dimensional. These three subspaces have an associated fitness
score of about 0.91. For methane PG-EA converges using $m=100$ chromosomes,
$\eta=0.4$, $\mu=0.01$ and 50 epochs. The Likelihood plot is represented
in panel (a) in the left part of figure \ref{fig:dendrogram_methane}.

As shown in Fig. \ref{fig:dendrogram_methane} (panel (b)), the 2-mode
approximation leads, in this case, to a poor subspace separation:
the best branch in the dendrogram is the colored (1,3,4)(2,5,6)(7)(8)(9)
with a score of 4.84, leading to a bigger error than PG-EA for the
Jacobian factorization.

\begin{figure}[h]
\centering
\includegraphics[clip,width=1.0\columnwidth]{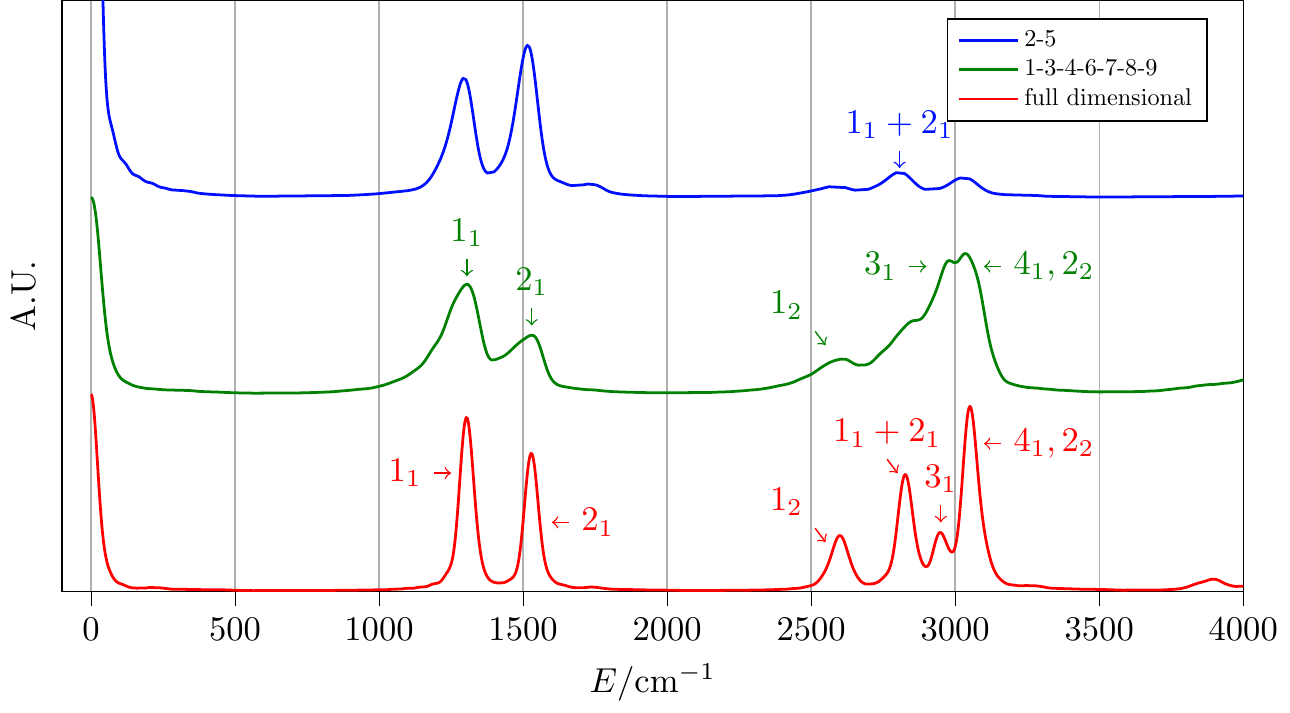}
\caption{Spectrum of methane: the full-dimensional spectrum is in red; the
reduced dimensionality spectra chosen according to PG-EA are in green
and blue.}
\label{fig:spectrum_methane} 
\end{figure}

Methane, in absence of a preliminary adiabatic switching sampling,\cite{Conte_Ceotto_ASSCIVR_2019}
is known to be characterized by highly chaotic dynamics, \cite{lee_taylor_PESch4_1995}
thus we employed 180000 trajectories to provide convergent results,
with a rejection rate of about 90\%, keeping nearly 2000 trajectories
per degree of freedom. Fig. \ref{fig:spectrum_methane} reports both
the full-dimensional spectrum (red continuous line) and the partial
dimensional ones (green and blue), according to the PG-EA vibrational
space sub-division found above. All spectroscopic features are properly
reproduced. Note that degenerate modes belonging to different subspaces
give rise to spectral lines at the same energy (see for instance $1_{1}$
and $2_{1}$ signals displayed in both subspace spectra in Fig. \ref{fig:spectrum_methane}).
In these cases, we consider more accurate the peaks appearing in the
largest subspace, as more mode interactions are taken into account,
even if frequencies of degenerate modes in different subspaces are
very similar and cannot be distinguished by naked eye. In Fig. \ref{fig:spectrum_methane}
we employ an incremental notation for the spectral features, so that
degenerate signals are collected together under the same label. For
a deeper insight, we report in Tab. \ref{tab:methane} the value in
wavenumbers of each spectral peak frequency. 
\begin{table*}[t]
\caption{Quantum frequencies of vibration of the methane molecule in $\mathrm{cm}^{-1}$
calculated on the PES by Lee, Martin and 
Taylor \cite{lee_taylor_PESch4_1995} using full-dimensional TA SCIVR,
DC-SCIVR based on PG-EA subspace partition, and discrete variable
representation calculations (Exact). MAE stands for Mean Absolute
Error, calculated using exact (MAE (Exact)) or full-dimensional semiclassical
values (MAE (TA SCIVR)) as reference.
In the fourth column, we report the DC-SCIVR frequencies obtained from the 
subdivision proposed in Ref.\citenum{DiLiberto_Ceotto_Jacobiano_2018} where a 
different approach for the Jacobi method was employed.
}
\label{tab:methane}
\begin{tabular}{
>{\raggedright}p{2.5cm}
>{\raggedright}p{2.5cm}
>{\raggedright}p{1.5cm}
>{\raggedright}p{2.5cm}
>{\raggedright}p{2.5cm}
>{\raggedright}p{4.0cm}}
Incremental label & 
Modes (symmetry)  & 
Exact \cite{Carter_Bowman_Methane_1999} & 
TA SCIVR &
DC SCIVR (1)(2,3)(4-9)\cite{DiLiberto_Ceotto_Jacobiano_2018} &
DC SCIVR PG-EA (2,5)(1,3,4,5-9) [sub]$^{a}$\tabularnewline
\hline 
$1_{1}$ & 1,2,3 $(F_{2})$ & 1313 & 1304 & 1287 & 1305 {[}G{]}\tabularnewline
$2_{1}$ & 4,5 $(E)$ & 1535 & 1529 & 1534 & 1530 {[}G{]}\tabularnewline
$1_{2}$ & 1,2,3 & 2624 & 2600 & 2562 & 2610 {[}G{]}\tabularnewline
$1_{1}+2_{1}$ & 1,2,3,4,5 & 2836 & 2827 &  & 2807 {[}B{]}\tabularnewline
$3_{1}$ & 6 $(A_{1})$ & 2949 & 2948 & 2960 & 2980 {[}G{]}\tabularnewline
$4_{1}$ & 7,8,9 $(F_{2})$ & 3053 & 3051 & 3044 & 3036 {[}G{]}\tabularnewline
$2_{2}$ & 4,5 & 3067 & 3051 & 3044 & 3036 {[}G{]}\tabularnewline
\hline 
\multicolumn{2}{l}{MAE (Exact)} &  & 9.6 & 22.0 & 19.3\tabularnewline
\hline 
\multicolumn{2}{l}{MAE (TA SCIVR)} &  &  & 14.3 & 13.4\tabularnewline
\end{tabular}

{$^{a}$subspace from which the wavenumber is taken: G for the 7-D
green one and B for 2-D blue one with reference to Figure \ref{fig:spectrum_methane}}
\end{table*}

In conclusion, PG-EA provides a subdivision of
the vibrational space appropriate for DC-SCIVR spectroscopic calculations
and we can move to apply it to larger systems where previous recipes
for vibrational space subdivision are impracticle.

\subsection{\emph{Trans}-N-Methylacetamide}

\label{sec:NMA} 
\begin{figure}[H]
\centering
\includegraphics[clip,width=0.5\textwidth]{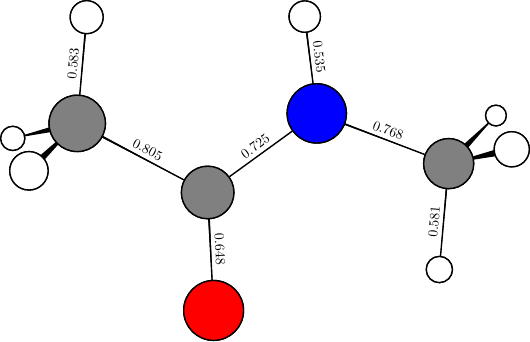}
\caption{\emph{trans}-N-Methylacetamide equilibrium geometry on the Potential
Energy Surface by Nandi, Qu, and Bowman.\cite{Qu_Bowman_FragmentationPES_2019}
Representative bond lengths of the equilibrium geometry are shown
in Ångström. The HNCO peptide bond is of fundamental importance for
the dynamics of peptides.}
\label{fig:NMApic} 
\end{figure}

Here we present the subspace optimization and the associated DC-SCIVR
spectroscopic calculations for the 30-mode \emph{trans}-N-Methylacetamide
(NMA) molecule represented in Fig. \ref{fig:NMApic}. NMA has been
studied thoroughly both computationally \cite{chen1995vibrational,kubelka2001ab,torii1998effects}
and experimentally \cite{ataka1984infrared,mayne1991resonance,chen1995vibrational,torii1998effects,triggs1992investigation},
as it is one of the simplest examples of a molecule featuring the
HNCO peptide bond. We use the full-PES by Nandi, Qu and Bowman\cite{Qu_Bowman_FragmentationPES_2019},
which has been designed both for \emph{cis-} and \emph{trans-} NMA
and that accounts for the three-fold symmetry of the methyl rotors.
The PES is permutationally invariant and was fitted to thousands of
ab initio calculated energies and gradients at B3LYP/cc-pVDZ level
of theory.\cite{Qu_Bowman_FragmentationPES_2019} In this case we
can compare our DC-SCIVR spectroscopic results with harmonic frequencies
and gas phase IR and Raman experimental values.\cite{ataka1984infrared}

The two methyl rotational frequencies, i.e. the two lowest-frequency
normal mode values, are not considered to be part of the vibrational
space. Thus, the dimensionality of the vibrational space we consider
is 28. We use a simulation time step of 5 a.u. The finite difference
displacement of the $i$-th normal mode for Hessian and gradients is
rescaled by $10^{-3}\sqrt{\max(\boldsymbol{\omega})/\bm{\omega}_{i}}$,\cite{Cazzaniga_Ceotto_skywalker_2020}
to account for different PES curvatures along each one of the normal
mode coordinates. We use the signal obtained from a single 30000 a.u.
long trajectory with initial conditions $(\sqrt{\boldsymbol{\omega}},\bm{q}_{eq})$.
We do not observe any conformational change from the \emph{trans}
to the \emph{cis} potential energy basin during our simulations.

\begin{figure}[h]
\centering
\includegraphics[width=1.0\columnwidth]{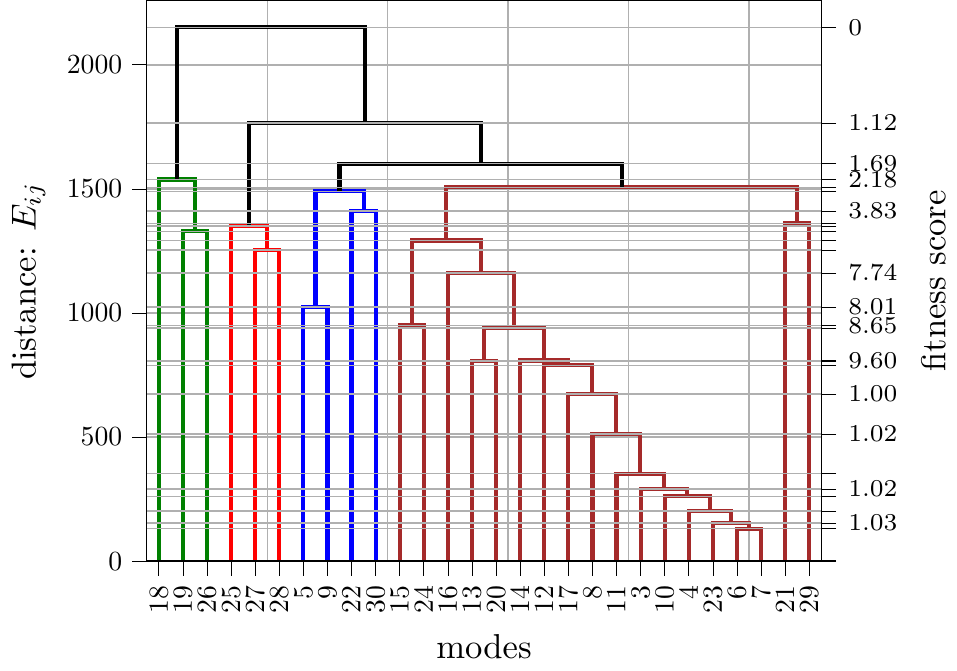}
\caption{2-mode interaction dendrogram of \emph{trans}-NMA. Four subspaces
are generated with a Jacobian factorization score equal to 2.18. As
an alternative, one might prefer the five-subspace option with an
error of 3.83.
We do not show the likelihood vs epochs plot, since the convergence of the 
whole population is not achieved nor desired. Furthermore computing the 
likelihood in Eq. \ref{eq:likelihood_cay}
for a 12 dimensional subspaces is not feasible as it requires the computation of
$12^{10}$ products.}
\label{fig:dendrogramNMA}
\end{figure}
We apply PG-EA, the 2-mode interaction method, and the Hessian method\cite{DiLiberto_Ceotto_Jacobiano_2018}
to divide the vibrational space into subspaces. The results are quite
different. For PG-EA, we run a thorough optimization using 10000 epochs,
$m=300$ chromosomes, $\mu=0.01$ and $\eta=0.4$, with the largest subspace
constraint set to $\bar{D}=15$ and we obtain the following three subspaces:
\textbf{A} = (3,5,7,10,11,12,15,21,23,26,28,30),
\textbf{B} = (4,9,14,16,17,18,19,22,24,25,27,29) and
\textbf{C} = (6,8,13,20).
The fitness score of the chromosome,
i.e. the score of the Jacobian factorization, is 1.96. When applying
the basic average Hessian criterion\cite{DiLiberto_Ceotto_Jacobiano_2018}
with a corse-graining parameter equal to \textbf{$8\cdot10^{-6}$},
we obtain the following subspaces:
$a=$(3,5,7,22),
$b=$(10,11,13,16,17,18,20,24,25,26,27,28,29),
$c=$(4),
$d=$(6),
$e=$(8),
$f=$(9),
$g=$(12),
$h=$(14),
$i=$(15),
$j=$(19),
$k=$(21),
$l=$(23),
$m=$(30),
which can be associated
to a fitness score equal to 4.49. Finally, we apply the 2-mode dendogram
approach and obtain the following subspaces:
$\alpha=$(3,4,6,7,8,10,11,12,13,14,15,16,17,20,21,23,24,29),
$\beta=$(5,9,22,30),
$\gamma=$(18,19,26), 
$\delta=$(25,27,28)
with
a fitness score of 2.18. The 2-mode interaction method produced the
dendrogram reported in Figure \ref{fig:dendrogramNMA}, which has
a slightly worse score than the PG-EA result. However, the presence
of an 18-dimensional subspace makes this subdivision not convenient
for Monte Carlo phase-space integration convergence. The Hessian method
is instead clearly penalized by the many 1-D subspaces found.
These considerations suggest that the PG-EA subdivision into three
subspaces is indeed the best choice.
\begin{figure*}[t]
\centering
\includegraphics[width=1\textwidth]{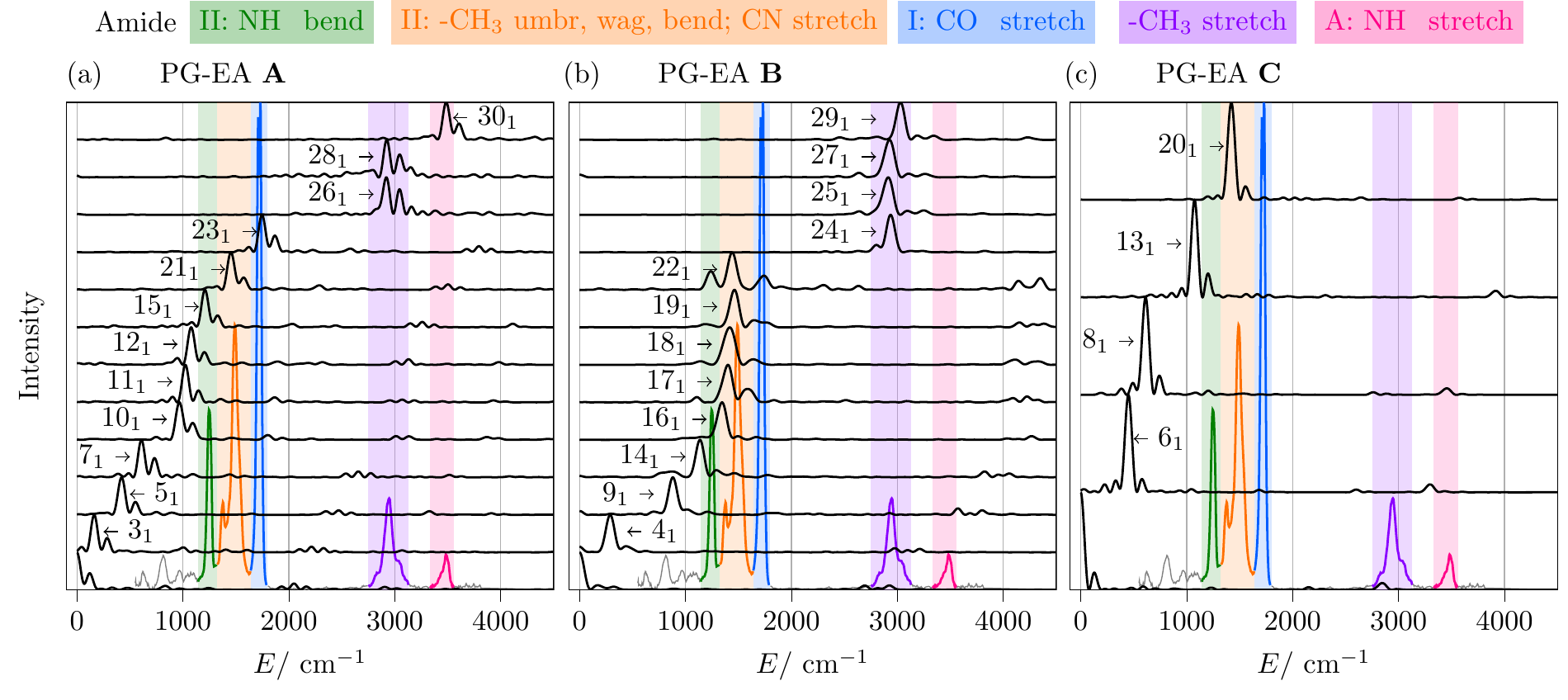}
\caption{Panels (a), (b), and (c) are the three partial spectra for the three
\textbf{A}, \textbf{B}, and \textbf{C} PG-EA subspaces. The colored
spectrum is the experimental IR spectrum. \cite{nist2020nmaspec}
The black continuous spectra are DC-SCIVR spectra calculated using
different $\epsilon_{j}$ value (see Eq.(\ref{eq:eps})) for each
normal mode of the subspace. Colored windows are for the different
spectroscopic regions as denominated by experimentalists. These windows
are commonly labeled Amide II (green and orange), Amide I (blue) and
Amide A (pink).
The purple window corresponds to \ce{CH3} stretching.
Amide III, IV, and V bands are located below $\sim1200\:\mathrm{cm}^{-1}$
and are very dependent on the side chains.}
\label{fig:spectrum_NMA_allsubs}
\end{figure*}
Based on PG-EA, we calculate the spectra reported in Fig. \ref{fig:spectrum_NMA_allsubs},
where the different spectral regions are highlighted by different
colors according to the experimentalists' denomination. Overall, the
DC-SCIVR spectra are reproducing well all the spectroscopic features
of this molecule. Actually there are more spectral features in the
DC-SCIVR simulation than there are in the experiment because DC SCIVR
calculates the power spectrum, which is made of all vibrational levels
(that we scale with respect to the zero-point energy), even those
associated to transitions which are not IR active. 
An IR spectrum simulation with related intensities would require 
calculation of the vibrational eigenfunctions.
This feature is not implemented yet in our divide-and-conquer
approach, but we are planning to do it soon.
In addition, according
to our simulations and referring to the experimentalists' denomination,
mode 23 is the only one responsible for the Amide I band, subspace
A contributes the most to amide III and A bands, while amide II is
mostly localized in subspace B.

\begin{table*}[t]
\caption{Vibrational fundamental frequencies for \emph{trans}-N-Methylacetamide
(NMA). The first column denominates the vibrational modes. In the
second column the fundamental frequencies in harmonic approximation
(HO) are reported. DC SCIVR fundamental frequency of vibrations are
obtained on the basis of subspace partition by means of PG-EA (third
column), Hessian method (Hess, fourth column), and 2-mode interaction
method (2-mode, fifth column). 
The results are sorted by increasing value of the harmonic frequencies 
and assigned by comparing the associated vibrational motion to the
corresponding experimental description.
Superscripts refer to the subspace
each mode belongs to. The sixth column reports the \emph{ab-initio}
harmonic frequencies at the MP2/aug-cc-pVTZ level of theory (HO/MP2)
.\cite{kaledin2007full} All data are compared by calculating the
Mean Absolute Error (MAE) with respect to the experimental values
(last column, Exp) by Ataka, Takeuchi and Tasumi.\cite{ataka1984infrared}}
\label{tab:NMA}
\setlength\extrarowheight{-7pt}
\begin{tabular}
{>
{\raggedleft}p{2.0cm}>
{\raggedleft}p{1.0cm}>
{\raggedleft}p{3.0cm}>
{\raggedleft}p{2.6cm}>
{\raggedleft}p{3.0cm}>
{\raggedleft}p{2.5cm}>
{\raggedleft}p{1.5cm}
}
\multicolumn{1}{c}{} & \multicolumn{4}{c}{Frequency $^{\mathrm{subID}}$ / $\mathrm{cm^{-1}}$} & \tabularnewline
\hline 
modes \# & HO & DC SCIVR PG-EA (3 subs) & DC SCIVR Hess (12 subs) & DC SCIVR 2-mode (4 subs) & HO/MP2\cite{kaledin2007full} & Exp \cite{ataka1984infrared}\tabularnewline
\hline 
3  & 150  & 163 $^{A}$ & 159 $^{a}$ & 156 $^{\alpha}$ & 151 & \tabularnewline
4  & 290  & 289 $^{B}$ & 285 $^{c}$ & 283 $^{\alpha}$ & 259 & 279\tabularnewline
5  & 393  & 421 $^{A}$ & 417 $^{a}$ & 416 $^{\beta}$ & 347 & 429\tabularnewline
6  & 433  & 451 $^{C}$ & 451 $^{d}$ & 448 $^{\alpha}$ & 423 & 439\tabularnewline
7  & 621  & 609 $^{A}$ & 608 $^{a}$ & 606 $^{\alpha}$ & 630 & 619\tabularnewline
8  & 629  & 613 $^{C}$ & 611 $^{e}$ & 612 $^{\alpha}$ & 633 & 658\tabularnewline
9  & 866  & 881 $^{B}$ & 875 $^{f}$ & 871 $^{\beta}$ & 883 & 857\tabularnewline
10 & 995  & 967 $^{A}$ & 970 $^{b}$ & 972 $^{\alpha}$ & 1003 & 980\tabularnewline
11 & 1038 & 1024 $^{A}$ & 1028 $^{b}$ & 1025 $^{\alpha}$ & 1058 & 1037\tabularnewline
12 & 1112 & 1078 $^{A}$ & 1080 $^{g}$ & 1080 $^{\alpha}$ & 1119 & 1089\tabularnewline
13 & 1132 & 1075 $^{C}$ & 1069 $^{b}$ & 1072 $^{\alpha}$ & 1169 & \tabularnewline
14 & 1166 & 1137 $^{B}$ & 1138 $^{h}$ & 1134 $^{\alpha}$ & 1195 & 1168\tabularnewline
15 & 1260 & 1208 $^{A}$ & 1210 $^{i}$ & 1205 $^{\alpha}$ & 1290 & 1266\tabularnewline
16 & 1391 & 1345 $^{B}$ & 1344 $^{b}$ & 1347 $^{\alpha}$ & 1402 & 1370\tabularnewline
17 & 1415 & 1401 $^{B}$ & 1388 $^{b}$ & 1388 $^{\alpha}$ & 1460 & 1419\tabularnewline
18 & 1434 & 1418 $^{B}$ & 1409 $^{b}$ & 1415 $^{\gamma}$ & 1487 & 1432\tabularnewline
19 & 1474 & 1462 $^{B}$ & 1463 $^{j}$ & 1461 $^{\gamma}$ & 1494 & 1446\tabularnewline
20 & 1485 & 1422 $^{C}$ & 1426 $^{b}$ & 1430 $^{\alpha}$ & 1499 & 1432\tabularnewline
21 & 1491 & 1453 $^{A}$ & 1453 $^{k}$ & 1453 $^{\alpha}$ & 1529 & 1472\tabularnewline
22 & 1550 & 1441 $^{B}$ & 1444 $^{a}$ & 1472 $^{\beta}$ & 1561 & 1511\tabularnewline
23 & 1772 & 1774 $^{A}$ & 1747 $^{l}$ & 1745 $^{\alpha}$ & 1749 & 1707\tabularnewline
24 & 3019 & 2936 $^{B}$ & 2934 $^{b}$ & 2920 $^{\alpha}$ & 3088 & 2915\tabularnewline
25 & 3040 & 2914 $^{B}$ & 2923 $^{b}$ & 2919 $^{\delta}$ & 3091 & 2958\tabularnewline
26 & 3072 & 2920 $^{A}$ & 2903 $^{b}$ & 2908 $^{\gamma}$ & 3165 & 2916\tabularnewline
27 & 3125 & 2926 $^{B}$ & 2933 $^{b}$ & 2924 $^{\delta}$ & 3188 & 3008\tabularnewline
28 & 3126 & 2924 $^{A}$ & 2912 $^{b}$ & 2912 $^{\delta}$ & 3188 & 3008\tabularnewline
29 & 3137 & 3030 $^{B}$ & 3044 $^{b}$ & 3037 $^{\alpha}$ & 3197 & 2973\tabularnewline
30 & 3630 & 3485 $^{A}$ & 3484 $^{m}$ & 3485 $^{\beta}$ & 3703 & 3498\tabularnewline
\hline 
MAE Exp & 47.9 & 30.0 & 29.7 & 27.5 & 78.5 & \tabularnewline
\end{tabular}

\end{table*}
All these subspace choices produce spectra with almost the same MAEs,
as reported in Tab. \ref{tab:NMA}. This suggests that the subspace
choice is flexible, as well as the choice of the subdivision criterion.
However, PG-EA is the method that minimizes the number of subspaces
and prevents from having many 1-D subspaces, which could result into a very
noisy and not resolved spectrum.\cite{Cazzaniga_Ceotto_skywalker_2020}

A closer inspection of the vibrational frequency values in Tab \ref{tab:NMA}
allows us to better understand the physical meaning of the MAE of
the different methods, in particular of the Harmonic versus the DC-SCIVR
one. In the case of the harmonic frequencies reported in the second
column of Tab. (\ref{tab:NMA}), 22 out of 26 vibrational frequencies
are higher than the experimental values. Thus, the MAE value of $48\:\text{cm}^{-1}$
is because of estimates by excess. In the case of the DC-SCIVR calculations
it is the other way around. For example, 20 PG-EA values are underestimating
the experimental frequencies and the MAE value of $\text{30\:cm}^{-1}$
is given by estimates by defect. Thus, the amount of anharmonic contribution
introduced by the DC-SCIVR calculations is on average per mode $\sim78\:\text{cm}^{-1}.$
This amount is comparable with the MAE under the column HO/MP2, where
harmonic frequencies are calculated at a higher level of \emph{ab
initio} theory than the DFT-B3LYP/cc-pVDZ one. These considerations
are suggesting that most probably, for this molecule, the discrepancies
with respect to the experimental values are mainly due to the DFT
level of \emph{ab initio} theory.
Conversely, only at a lower degree the inaccuracy can be related to the 
semiclassical approximation or the quality of the potential energy surface 
fitting, as previously shown on other systems.\cite{conte2020sensitivity}
Unfortunately, a DC-SCIVR simulation at MP2/aug-cc-pVTZ level of \emph{ab
initio} theory is out of reach at time of writing due to its computational
burden.

\section{Summary and Conclusions\label{sec:Conclusions}}

We have presented a machine learning algorithm based on a Probability
Graph representation and an Evolutionary Algorithm procedure. The
algorithm is able to find the best subdivision of the full-dimensional
vibrational space into subspaces for model systems in which the best
subspace division is known. Our approach is able to preserve Liouville's
theorem for each subspace as much as possible and for a given maximum
dimensionality of the subspaces. We proved that the clustering provided
by PE-GA is indeed one of the possible solutions that minimize the
energy exchange between subspaces during the vibrational dynamics,
and thus the most convenient for DC-SCIVR and spectroscopic calculations
in general. As an alternative, we have proposed a 2-mode coupling
scheme which is less computational intense but also less accurate.
Application to the DC-SCIVR power spectrum calculation of \emph{trans}
N-Methylacetamide is made manageable under Liouville's criterion restrictions
only by means of these algorithms.
The calculation of the DC-SCIVR power spectrum of trans-NMA with subspace 
division selected with Liouville's criterion is manageable only employing
these two algorithms.

The choice of the PG-EA parameters is arbitrary to some extent. As
a matter of fact, the method is bound to look for new solutions at
every epoch and hence it will get the global optimum, eventually.
However, a sensible choice of the number of chromosomes and the mutation
probability may significantly enhance the optimization. Assuming that
the number of epochs is fixed, increasing the number of (elite) chromosomes
means that the population evolves more slowly, enhancing the chances
of eventually hitting the global optimum. However, as the evaluation
of the fitness function is the most expensive step, a large population
requires significantly more computational time. Conversely, using
a small population means a fast evolution and therefore it is likely
to obtain a fast local minimum. We suggest that a sensible choice
of the mutation probability is in the interval {[}0.001, 0.2{]}. This
parameter makes sure that the algorithm does not get stuck in a local
minimum, even if the whole population is homogeneous. It becomes less
and less important as the pool of possible solutions and the number
of elite chromosomes increase. 

There are several quantum methods that can take advantage from partitioning
the nuclear vibrational degrees of freedom. Clearly, dividing the
vibrational space into putative independent subspaces is an approximation.
However, if this subdivision is performed according to Jacobi's criterion,
it may turn out not to be a rough approximation, especially for high
dimensional and loosely connected systems. We believe that, at the
affordable cost of a single adiabatic classical trajectory with Hessian
calculation, the PG-EA algorithm can be useful to assist any method
that has to deal with increasing computational costs with system dimensionality
but that is able to perform accurate spectroscopic calculations for
each subspace independently. This may be the case, for example, of
the Local Mode variant of Multimode\cite{wang_bowman_ringhexamerLMM_2012,liu_bowman_LMMice_2012,wang_bowman_LMMhexamer_2013}
or other semiclassical wavepacket propagation methods developed by
other groups.\cite{Begusic_Vanicek_Electronicspectra_2018,Patoz_Vanicek_Photoabs_Benzene_2018,Wehrle_Vanicek_NH3_2015}

The work we have presented provides a rigorous rationalization of
the simplification of a larger dimensional problem into a set of lower
dimensional ones. The examples illustrated in the paper demonstrate
that reliable spectroscopic results are obtained if a rigorous strategy
is employed to get to a reasonable subspace partition while a non-educated,
unwise choice of subspaces may lead to inaccurate or unreliable results.
Our algorithms might serve as a powerful tool for advancing the computational
spectroscopy of large molecules.

\begin{acknowledgments}
Authors acknowledge financial support from the European Research Council
(Grant Agreement No. (647107)---SEMICOMPLEX---ERC-2014-CoG) under
the European Union’s Horizon 2020 research and innovation programme,
and from the Italian Ministery of Education, University, and Research
(MIUR) (FARE programme R16KN7XBRB- project QURE).
\end{acknowledgments}

\section*{AIP PUBLISHING DATA SHARING POLICY}
The data that support the findings of this study are available from the 
corresponding author upon reasonable request.

\bibliography{bibliography}

\end{document}